\documentclass[
  aps,
  prd,
  reprint,
  nofootinbib,
  superscriptaddress,
  longbibliography
]{revtex4-2}

\usepackage{amsmath}
\usepackage{amssymb}
\usepackage{graphicx}
\usepackage{bm}
\usepackage{xcolor}
\usepackage[colorlinks=true,linkcolor=blue,citecolor=blue,urlcolor=blue]{hyperref}

\newcommand{\Mpl}{M_{\rm Pl}}

\begin{document}

\title{Warm inflation beyond the Markovian limit}

\author{Mayukh R. Gangopadhyay}
\email{mayukhraj.g@vit.ac.in}
\affiliation{Department of Physics, School of Advanced Sciences, Vellore Institute of Technology (VIT), Chennai Campus, Chennai, 600127, Tamil Nadu, India}

\author{Nilanjana Kumar}
\email{nilanjanakumar@care.edu.in}
\affiliation{Department of Physics, Chettinad Institute of Technology, Chettinad Academy of Research and Education, Chennai, 603103, Tamil Nadu, India}

\begin{abstract}
Warm inflation is commonly studied under the assumption that the stochastic
force sourcing inflaton fluctuations is Markovian. Realistic thermal systems,
however, possess finite relaxation times and can therefore generate colored
noise with non-zero correlation time. In this work we investigate warm
inflation beyond the Markovian limit and determine how finite correlation time
modifies the primordial scalar power spectrum. We show that memory effects
suppress the scalar spectrum relative to the standard white-noise result and
derive a simple expression for this correction in terms of the background
thermal dynamics. In particular, we relate the size of the non-Markovian effect
directly to the thermal ratio between the bath temperature and the Hubble
scale, thereby establishing a transparent link between warm-inflation background
quantities and the validity of the Markovian approximation. We also derive the
corresponding modification of the tensor-to-scalar ratio and the induced shifts
in the scalar spectral index and the running of the scalar spectral index. Our
results provide a simple and practical diagnostic for identifying when finite
correlation-time effects become relevant in warm-inflation model building.
\end{abstract}

\maketitle

\section{Introduction}

Cosmic inflation remains the leading paradigm for explaining the large-scale
homogeneity, isotropy, and near-flatness of the Universe, as well as the origin
of primordial perturbations that later seeded structure formation~\cite{Guth1981,Linde1982,Albrecht1982,Linde1983,Linde1990,Lyth1999,Baumann2009,Weinberg2008,Mukhanov2005}.
In the standard cold inflation scenario the Universe undergoes accelerated
expansion in a nearly vacuum-dominated state and reheating occurs only after
inflation ends. Warm inflation offers a conceptually different realization in
which dissipative interactions between the inflaton and additional fields
continuously produce radiation during inflation~\cite{Berera1995,BereraFang1995,Berera2000,BereraMossRamos2009}.

The presence of radiation during inflation changes the structure of the
perturbation problem in an essential way. In warm inflation the inflaton evolves
in a thermal environment, and thermal fluctuations may contribute directly to
the curvature perturbation spectrum~\cite{TaylorBerera2000,deOliveiraJoras2001,HallMossBerera2004,MossXiong2006,MossXiong2007,GrahamMoss2009,BasteroGil2011,BasteroGilCMB2018,BasteroGilNonGauss2014}.
The perturbation dynamics is therefore naturally described in stochastic terms:
the inflaton fluctuation obeys a Langevin-type equation containing both a
dissipative term and a stochastic force. In most analyses this stochastic force
is assumed to be Markovian, corresponding to white noise with vanishing
correlation time. That approximation is sensible when the microscopic thermal
relaxation time is much shorter than the response timescale of the inflaton
mode.

Realistic thermal systems, however, are generally not strictly Markovian.
Interactions in a plasma occur over finite timescales and may generate memory
effects. From the viewpoint of nonequilibrium statistical mechanics, the
white-noise approximation is only a limiting case of a more general stochastic
dynamics in which the noise possesses finite correlation
time~\cite{Kubo1966,Mori1965,Zwanzig1961,Zwanzig2001,Gardiner2004,VanKampen2007,Risken1989,Fox1978,Hanggi1995}.
Similar non-Markovian structures arise in open quantum systems and
nonequilibrium field theory~\cite{CalzettaHu1988,HuPazZhang1992,CalzettaHu2008,Boyanovsky1996,Boyanovsky1997}.
In cosmology, stochastic approaches have also been central in the study of
inflationary fluctuations~\cite{Starobinsky1986,StarobinskyYokoyama1994,VenninStarobinsky2015,PattisonVennin2019,GrainVennin2017}.
These observations motivate a systematic investigation of finite
correlation-time effects in warm inflation.

The purpose of the present work is to analyze the consequences of colored noise
for the scalar power spectrum in warm inflation and to do so in a way that
remains closely tied to the background dynamics.

Several prior works have studied related aspects of non-Markovian or
frequency-dependent effects in warm and stochastic inflation. Ramos \&
da~Silva~\cite{RamosDaSilva2013} computed the power spectrum including both
quantum and thermal noise contributions using the stochastic inflation
approach, but treated the noise as strictly Markovian throughout.
Bastero-Gil et al.~\cite{BasteroGil2011} derived dissipation coefficients from
quantum field interactions, and the associated frequency-dependent structure of
$\Upsilon(\omega)$ is complementary to our treatment of the noise-correlation
side. The stochastic inflation framework of
Starobinsky--Yokoyama~\cite{Starobinsky1986,StarobinskyYokoyama1994} and its
extensions~\cite{VenninStarobinsky2015,GrainVennin2017,AndoVennin2021} address
quantum diffusion in de~Sitter but assume Markovian (white) noise throughout. In
the open quantum systems literature, the influence functional approach of
Calzetta \& Hu~\cite{CalzettaHu1988,CalzettaHu2008} generates memory kernels
from first principles; our Ornstein--Uhlenbeck parametrization is the minimal
effective model consistent with that framework in the classical stochastic
limit. For recent reviews of warm inflation developments
see~\cite{KamaliMotaharfarRamos2023,BasteroGilEFT2021}.

We show that the effect of finite correlation time can be encoded in the
dimensionless parameter
\begin{equation}
\eta_c = (3H + \Upsilon)\tau_c = 3H(1+Q)\tau_c ,
\label{eq:etac-def}
\end{equation}
where $Q = \Upsilon/(3H)$ is the dissipative ratio and $\tau_c$ is the
correlation time of the stochastic force. If $\eta_c \ll 1$, the stochastic
force decorrelates rapidly and the standard Markovian limit is recovered. If
$\eta_c \gtrsim 1$, memory effects become relevant and the scalar spectrum is
suppressed relative to the standard warm inflation result.

A key result of this paper is that $\eta_c$ can be expressed directly in terms of
the thermal ratio $T/H$, which itself follows from the warm-inflation background
equations. This establishes the pipeline
\begin{equation*}
(H, V', Q) \;\longrightarrow\; T/H \;\longrightarrow\; \eta_c(Q)
\;\longrightarrow\; \delta(Q),
\end{equation*}
where
\begin{equation*}
\delta(Q) \equiv \frac{P_{\rm col}}{P_{\rm white}}
\end{equation*}
is the scalar suppression factor. Once $\delta(Q)$ is known, the modification of
the tensor-to-scalar ratio and the induced shifts in $n_s$ and $\alpha_s$ follow
immediately.

A second important point is interpretational. The present work is not merely a
calculation of a correction to the scalar spectrum. It also provides a criterion
for the validity of the Markovian approximation used throughout much of the
warm-inflation literature. Indeed, since
\begin{equation*}
\eta_c = \frac{3(1+Q)}{\alpha(T/H)},
\end{equation*}
the standard white-noise treatment is valid only when
\begin{equation*}
\alpha(T/H) \gg 3(1+Q).
\end{equation*}
Conversely, when
\begin{equation*}
\alpha(T/H) \lesssim 3(1+Q),
\end{equation*}
finite correlation-time effects can no longer be ignored.

The paper is organized as follows. In Sec.~\ref{sec:background} we review the
background equations of warm inflation and derive the explicit expression for
$T/H$. In Sec.~\ref{sec:stochastic} we formulate the fluctuation problem beyond
the Markovian limit and derive in detail the relation between the colored and
white scalar spectra. In Sec.~\ref{sec:observables} we discuss the implications
for cosmological observables. In Sec.~\ref{sec:regimes} we analyze the physical
regimes of validity of the Markovian approximation and give a microphysical
interpretation of the correlation-time parameter. In Sec.~\ref{sec:numerical} we
summarize the numerical outputs and the associated plots. We conclude in
Sec.~\ref{sec:conclusions}. The Appendix contains a compact
Ornstein--Uhlenbeck derivation of the suppression factor.

\section{Warm inflation background and the thermal ratio \texorpdfstring{$T/H$}{T/H}}
\label{sec:background}

The homogeneous inflaton field obeys
\begin{equation}
\ddot{\phi} + 3H\dot{\phi} + V'(\phi) = -\Upsilon\dot{\phi},
\label{eq:eom}
\end{equation}
where $\Upsilon$ is the dissipative coefficient. The standard dissipative ratio
is defined by
\begin{equation}
Q \equiv \frac{\Upsilon}{3H}.
\label{eq:Qdef}
\end{equation}
In the slow-roll regime Eq.~\eqref{eq:eom} reduces to
\begin{equation}
3H(1+Q)\dot{\phi} \simeq -V'.
\label{eq:slowroll}
\end{equation}
The radiation bath evolves according to
\begin{equation}
\dot{\rho}_R + 4H\rho_R = \Upsilon\dot{\phi}^2.
\label{eq:radiation}
\end{equation}
During warm inflation one usually works in the quasi-steady regime in which
$\dot{\rho}_R$ is subleading, giving
\begin{equation}
4H\rho_R \simeq \Upsilon\dot{\phi}^2.
\label{eq:quasisteady}
\end{equation}
Using Eq.~\eqref{eq:Qdef}, this becomes
\begin{equation}
\rho_R = \frac{\Upsilon}{4H}\dot{\phi}^2 = \frac{3}{4}\,Q\,\dot{\phi}^2.
\label{eq:rhoR1}
\end{equation}
Substituting the slow-roll velocity from Eq.~\eqref{eq:slowroll},
\begin{equation}
\dot{\phi}^2 = \frac{V'^2}{9H^2(1+Q)^2},
\label{eq:phidot2}
\end{equation}
one obtains
\begin{equation}
\rho_R = \frac{Q\,V'^2}{12H^2(1+Q)^2}.
\label{eq:rhoR2}
\end{equation}
The temperature of the thermal bath is related to the radiation density by
\begin{equation}
\rho_R = \frac{\pi^2}{30}\,g_*\,T^4,
\label{eq:rhoR-T}
\end{equation}
with $g_*$ the effective number of relativistic degrees of freedom. Combining
Eqs.~\eqref{eq:rhoR2} and~\eqref{eq:rhoR-T} gives
\begin{equation}
T^4 = \frac{30}{\pi^2 g_*}\,\frac{Q\,V'^2}{12H^2(1+Q)^2}.
\label{eq:T4}
\end{equation}
Dividing by $H^4$ yields the explicit thermal ratio
\begin{equation}
\frac{T}{H} = \left[\frac{30}{\pi^2 g_*}\,
\frac{Q\,V'^2}{12H^6(1+Q)^2}\right]^{1/4}.
\label{eq:ToverH}
\end{equation}
Equation~\eqref{eq:ToverH} makes the dependence on the dissipative ratio
completely explicit. For fixed $H$, $V'$, and $g_*$, the nontrivial $Q$
dependence is controlled by $Q/(1+Q)^2$. Consequently,
\begin{align}
Q \ll 1 &: \quad \frac{T}{H} \propto Q^{1/4}, \label{eq:weak}\\
Q \gg 1 &: \quad \frac{T}{H} \propto Q^{-1/4}. \label{eq:strong}
\end{align}
Thus the thermal ratio typically increases in the weak dissipative regime, peaks
around $Q \sim \mathcal{O}(1)$, and decreases slowly for strong dissipation.

\section{Stochastic dynamics beyond the Markovian limit}
\label{sec:stochastic}

\subsection{Langevin equation}

The Fourier modes of the inflaton perturbation satisfy the stochastic equation
\begin{equation}
\ddot{\delta\phi}_k + (3H + \Upsilon)\dot{\delta\phi}_k
+ \frac{k^2}{a^2}\,\delta\phi_k = \xi_k,
\label{eq:langevin}
\end{equation}
where $\xi_k$ is the stochastic force sourced by interactions with the thermal
bath. In the Markovian approximation one takes $\xi_k$ to be white noise,
\begin{equation}
\langle\xi(t)\xi(t')\rangle \propto \delta(t-t').
\label{eq:white}
\end{equation}
To go beyond this limit we assume that the force has finite correlation time
$\tau_c$. The simplest and most useful model is exponentially correlated noise,
\begin{equation}
\langle\xi(t)\xi(t')\rangle = D\,e^{-|t-t'|/\tau_c},
\label{eq:OUcorr}
\end{equation}
which is the standard Ornstein--Uhlenbeck kernel. A simple realization of
colored noise is provided by the Ornstein--Uhlenbeck process
\begin{equation}
\dot{\xi} = -\frac{\xi}{\tau_c} + \eta(t),
\label{eq:OUprocess}
\end{equation}
where $\eta(t)$ is white noise. Solving this process yields the exponential
correlator given in~\eqref{eq:OUcorr}. The Ornstein--Uhlenbeck (OU)\footnote{For
the remainder of the paper we will use the short form OU to refer to
Ornstein--Uhlenbeck.} kernel is the natural minimal choice for the following
reason. By Doob's theorem, the OU process is the unique stationary Gaussian
Markov process with finite correlation time~\cite{Gardiner2004}. In a weakly
coupled thermal plasma the bath relaxation is dominated by a single rate
$\Gamma_{\rm th} \sim g^2 T$, so the noise decorrelates exponentially and the OU
kernel is the appropriate effective description. Power-law or
stretched-exponential kernels arise in systems with a broad spectrum of
relaxation rates (e.g.\ glassy or fractal media), which is not the physical
situation in warm inflation. Moreover, for any kernel with the same correlation
time $\tau_c$, the suppression factor takes the form
$\delta = 1/(1 + \eta_c^2 \cdot F)$ where $F$ is a kernel-dependent form factor
of order unity; for the OU kernel $F = 1$ exactly, so the result is robust at the
order-of-magnitude level to the specific kernel shape~\cite{Fox1978,Hanggi1995}.

\subsection{Colored-noise parameter}

The deterministic part of Eq.~\eqref{eq:langevin} contains the damping timescale
\begin{equation}
\tau_d = (3H + \Upsilon)^{-1} = \frac{1}{3H(1+Q)}.
\label{eq:taud}
\end{equation}
It is therefore natural to define the dimensionless colored-noise parameter
\begin{equation}
\eta_c \equiv (3H + \Upsilon)\tau_c = 3H(1+Q)\tau_c.
\label{eq:etac}
\end{equation}
This parameter measures the ratio of the noise correlation time to the
relaxation time of the inflaton perturbation.

To connect the correlation time to thermal physics we parametrize
\begin{equation}
\tau_c = \frac{1}{\alpha T},
\label{eq:tauc}
\end{equation}
where $\alpha$ is an effective inverse correlation-time parameter for the
inflaton-sourcing sector. Microphysically, the thermal relaxation rate in a
weakly coupled plasma scales as $\Gamma_{\rm th} \sim g^2 T$ (from
hard-thermal-loop resummation and Landau damping), so
$\tau_c \sim 1/(g^2 T)$ and one identifies $\alpha \sim g^2$. For coupling
constants in the range $g = 0.05$--$0.5$, this gives
$\alpha \in [0.0025, 0.25]$. Combining Eq.~\eqref{eq:tauc} with the background
result~\eqref{eq:ToverH}, one obtains
\begin{equation}
\eta_c(Q) = \frac{3(1+Q)}{\alpha(T/H)}.
\label{eq:etacQ}
\end{equation}
For notational compactness we write $A \equiv \alpha(T/H)$, so
$\eta_c = 3(1+Q)/A$.

\subsection{Detailed derivation of the scalar suppression factor}

Equation~\eqref{eq:langevin} is linear, so the formal solution may be written in
terms of the retarded Green function $G_k(t,t')$:
\begin{equation}
\delta\phi_k(t) = \int^t dt'\, G_k(t,t')\,\xi_k(t').
\label{eq:greensoln}
\end{equation}
The fluctuation variance then follows as
\begin{equation}
\langle|\delta\phi_k(t)|^2\rangle =
\int^t dt_1 \int^t dt_2\, G_k(t,t_1) G_k(t,t_2)\,
\langle\xi(t_1)\xi(t_2)\rangle.
\label{eq:variance}
\end{equation}
In the white-noise limit the correlator is proportional to
$\delta(t_1 - t_2)$, and the double integral reduces to the familiar Markovian
result. For exponentially correlated noise, by contrast, the source remains
correlated over a finite interval of width $\tau_c$.

In the present problem the relevant response timescale is the damping time
$\tau_d = (3H + \Upsilon)^{-1}$. If the response is dominated by this single
effective timescale, the influence of the exponential kernel can be estimated
analytically. This single-timescale approximation is valid when:
(i) the mode is superhorizon ($k/aH \ll 1$), so the $k^2/a^2$ term in
Eq.~\eqref{eq:langevin} is subdominant relative to $(3H+\Upsilon)^2$;
(ii) the potential curvature contributes only at slow-roll order
($\eta_V H^2 \ll (3H+\Upsilon)^2$); and (iii) all timescales can be approximated
by $\tau_d$. Since the primordial power spectrum is evaluated at horizon
crossing $k = aH$, where $k^2/a^2 = H^2 \ll (3H+\Upsilon)^2$ for
$Q \gtrsim \mathcal{O}(1)$, condition (i) is self-consistently satisfied. In the
transitional regime $\eta_c \sim 1$, the approximation introduces at most an
$\mathcal{O}(1)$ error in $\delta$, which does not affect the qualitative regime
boundaries. In this single-timescale approximation one finds
\begin{equation}
\langle|\delta\phi_k|^2\rangle_{\rm col} =
\frac{1}{1+\eta_c^2}\,\langle|\delta\phi_k|^2\rangle_{\rm white}.
\label{eq:varsupp}
\end{equation}
Since the curvature perturbation is proportional to the inflaton fluctuation,
the same suppression factor carries directly to the scalar power spectrum:
\begin{equation}
P_{\rm col}(k) = \delta(Q)\,P_{\rm white}(k),
\qquad
\delta(Q) \equiv \frac{1}{1+\eta_c(Q)^2}.
\label{eq:Pcol}
\end{equation}
This single equation encodes the central result of the present work. The
suppression factor $\delta(Q)$ is fully determined by the background
warm-inflation quantities $(H, V', Q, g_*)$ through the pipeline
$(H, V', Q) \to T/H \to \eta_c(Q) \to \delta(Q)$. This relation provides a direct
diagnostic for the validity of the Markovian approximation in warm inflation.

\section{Consequences for cosmological observables}
\label{sec:observables}

Since tensor perturbations remain vacuum generated and do not couple directly to
the thermal stochastic force at leading order, the tensor spectrum is unchanged
while the scalar spectrum is suppressed. The tensor-to-scalar ratio therefore
becomes
\begin{equation}
r_{\rm col} = \frac{r_{\rm white}}{\delta(Q)}.
\label{eq:rcol}
\end{equation}
Equivalently,
\begin{equation}
\frac{r_{\rm col}}{r_{\rm white}} = \frac{1}{\delta(Q)}.
\label{eq:rratio}
\end{equation}
The scalar spectral index is defined by
\begin{equation}
n_s - 1 \equiv \frac{d\ln P_\zeta}{d\ln k}.
\label{eq:nsdef}
\end{equation}
Using $P^{\rm col}_\zeta = \delta\, P^{\rm white}_\zeta$, we have
\begin{equation}
n^{\rm col}_s - 1 = (n^{\rm white}_s - 1) + \frac{d\ln\delta}{d\ln k}.
\label{eq:nscol}
\end{equation}
Hence the colored-noise shift in the scalar tilt is
\begin{equation}
\Delta n_s \equiv n^{\rm col}_s - n^{\rm white}_s
= \frac{d\ln\delta}{d\ln k}.
\label{eq:Deltans}
\end{equation}
Similarly, the running is defined by
\begin{equation}
\alpha_s \equiv \frac{dn_s}{d\ln k},
\label{eq:alphadef}
\end{equation}
so the colored-noise contribution is
\begin{equation}
\Delta\alpha_s \equiv \alpha^{\rm col}_s - \alpha^{\rm white}_s
= \frac{d^2\ln\delta}{d(\ln k)^2}.
\label{eq:Deltaalpha}
\end{equation}
To evaluate these derivatives in a simple and plot-ready way we introduce a mild
scale dependence
\begin{equation}
Q(k) = Q_*\left(\frac{k}{k_*}\right)^q,
\label{eq:Qk}
\end{equation}
which implies
\begin{equation}
\frac{d\ln Q}{d\ln k} = q.
\label{eq:dlnQ}
\end{equation}
By the chain rule,
\begin{equation}
\Delta n_s = q\,\frac{d\ln\delta}{d\ln Q},
\qquad
\Delta\alpha_s = q^2\,\frac{d^2\ln\delta}{d(\ln Q)^2}.
\label{eq:chainrule}
\end{equation}
In principle $q = d\ln Q/d\ln k$ is derivable from the underlying model via the
slow-roll equations; for the Warm Little Inflaton quartic potential one finds
$q \sim 0.01$--$0.05$ at typical horizon-crossing field values. The
$(q,\alpha)$ degeneracy can be partially broken because $q$ shifts both $n_s$ and
$\alpha_s$ simultaneously, while $\alpha$ shifts primarily $n_s$; the running
$\Delta\alpha_s$ therefore provides an independent handle on $q$ alone.

\begin{figure}[t]
  \centering
  \includegraphics[width=\columnwidth]{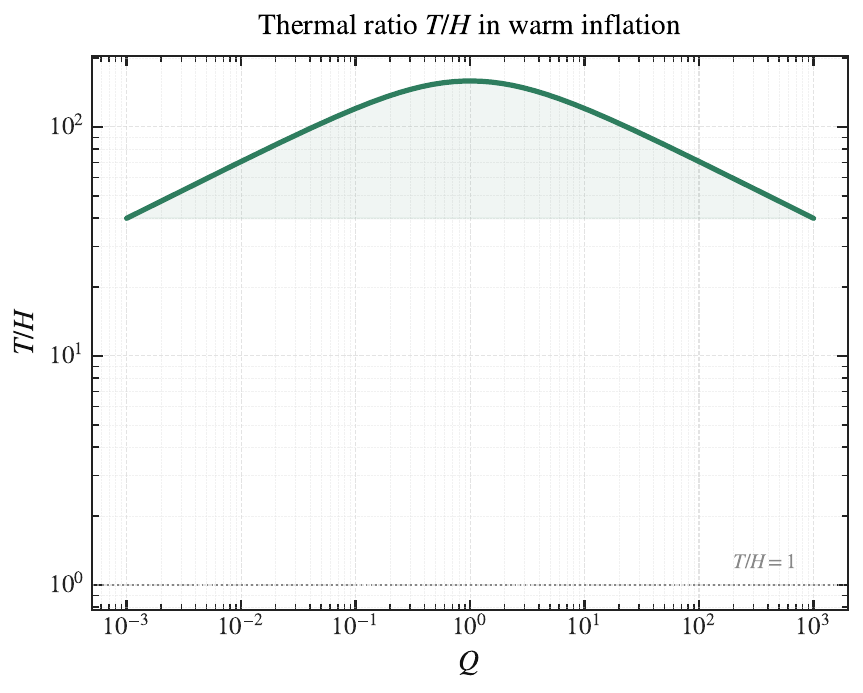}
  \caption{Thermal ratio $T/H$ as a function of the dissipative ratio $Q$
  obtained from Eq.~\eqref{eq:ToverH}.}
  \label{fig:ToverH}
\end{figure}

\begin{figure}[t]
  \centering
  \includegraphics[width=\columnwidth]{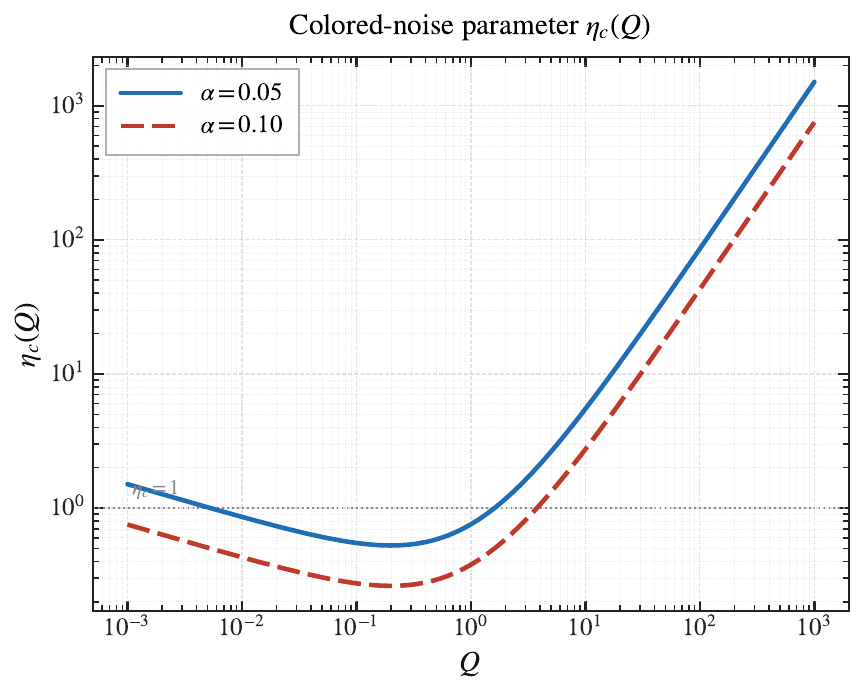}
  \caption{Colored-noise parameter $\eta_c(Q)$ computed from
  Eq.~\eqref{eq:etacQ}.}
  \label{fig:etac}
\end{figure}

\begin{figure}[t]
  \centering
  \includegraphics[width=\columnwidth]{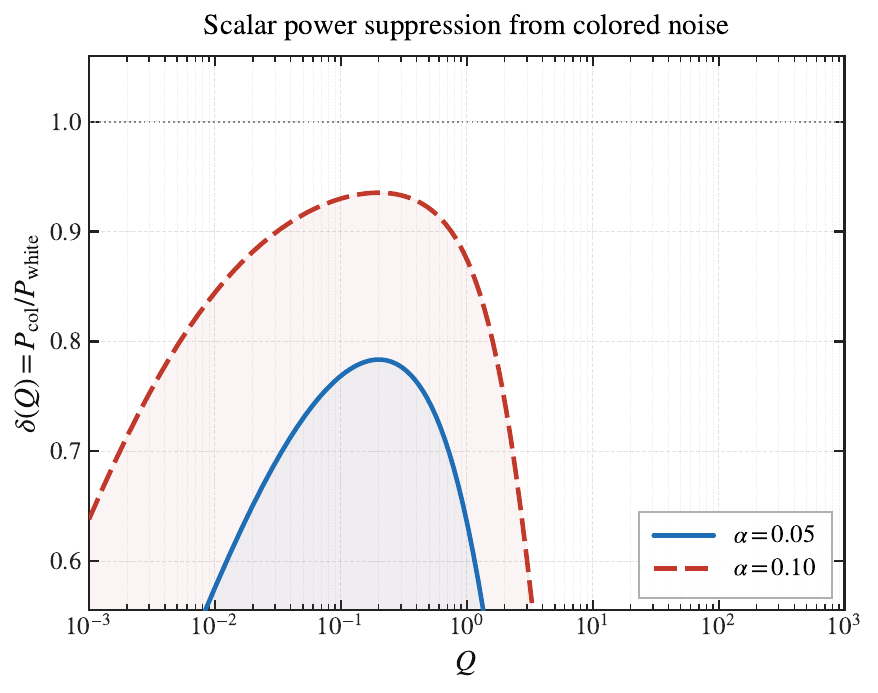}
  \caption{Scalar suppression ratio $P_{\rm col}/P_{\rm white} = \delta(Q)$
  obtained from Eq.~\eqref{eq:Pcol}.}
  \label{fig:delta}
\end{figure}

\begin{figure}[t]
  \centering
  \includegraphics[width=\columnwidth]{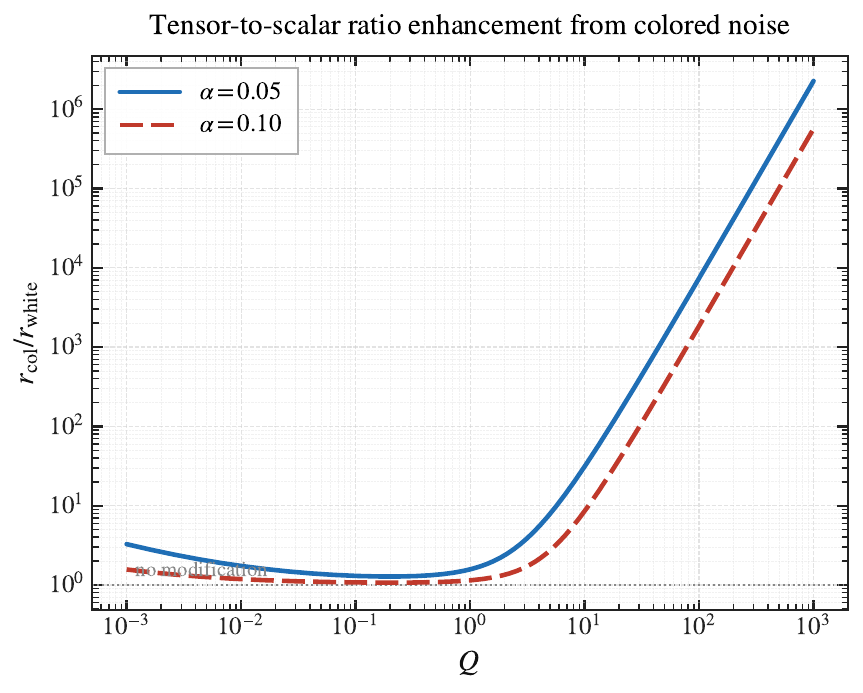}
  \caption{Enhancement of the tensor-to-scalar ratio due to colored noise,
  $r_{\rm col}/r_{\rm white} = 1/\delta(Q)$.}
  \label{fig:rratio}
\end{figure}

\begin{figure}[t]
  \centering
  \includegraphics[width=\columnwidth]{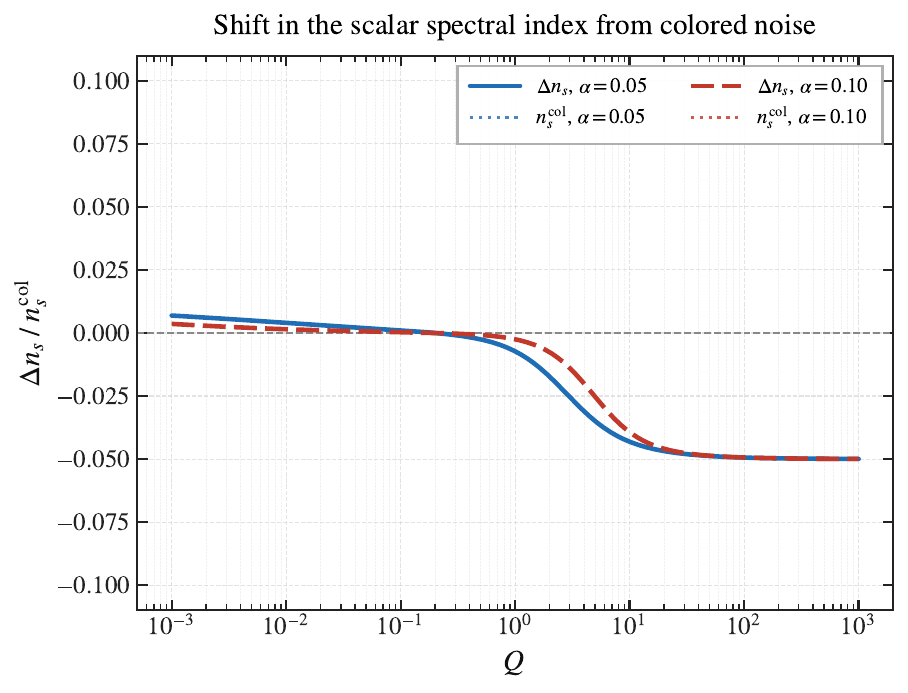}
  \caption{Colored-noise contribution to the spectral index,
  $\Delta n_s(Q) = d\ln\delta/d\ln k$, evaluated using the mild scale dependence
  of Eq.~\eqref{eq:Qk}.}
  \label{fig:Deltans}
\end{figure}

\begin{figure}[t]
  \centering
  \includegraphics[width=\columnwidth]{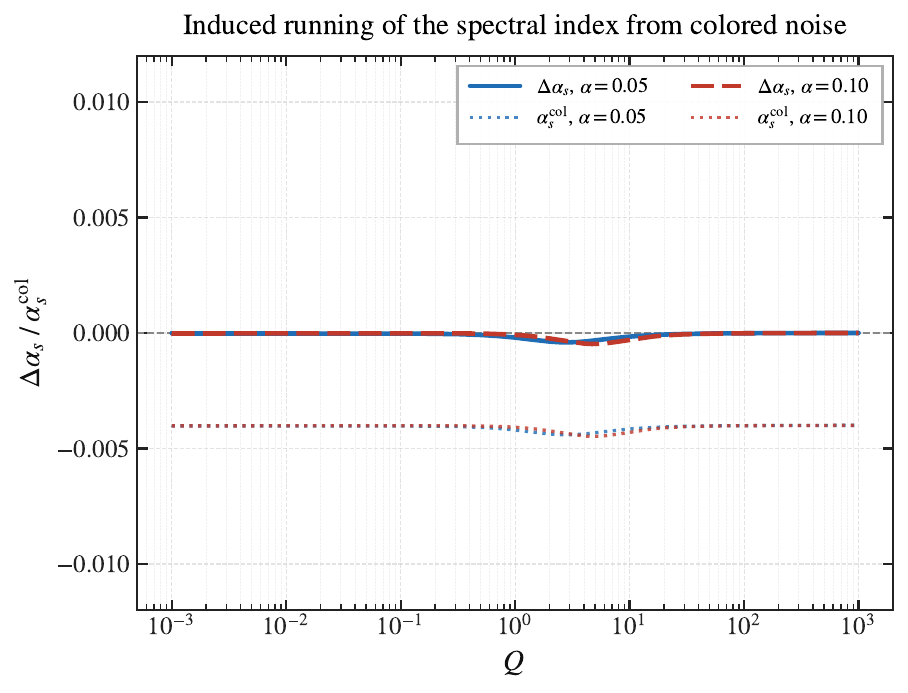}
  \caption{Colored-noise contribution to the running,
  $\Delta\alpha_s(Q) = d^2\ln\delta/d(\ln k)^2$.}
  \label{fig:Deltaalpha}
\end{figure}

\section{Regimes of validity and physical interpretation}
\label{sec:regimes}

Since $\eta_c = 3(1+Q)/[\alpha(T/H)]$ (see Eq.~\eqref{eq:etacQ}), the physics
naturally splits into three regimes.

In the deep Markovian regime,
\begin{equation}
\eta_c \ll 1 \quad\Longleftrightarrow\quad \alpha(T/H) \gg 3(1+Q),
\label{eq:deepmarkov}
\end{equation}
the correlation time of the stochastic force is much shorter than the response
time of the inflaton mode, and the white-noise approximation is valid.

In the transitional regime,
\begin{equation}
\eta_c \sim 1,
\label{eq:transitional}
\end{equation}
the correlation time becomes comparable to the inflaton relaxation time. In this
regime the Markovian approximation begins to fail and the scalar spectrum starts
to deviate appreciably from the standard result.

In the strong non-Markovian regime,
\begin{equation}
\eta_c \gg 1,
\label{eq:strongNM}
\end{equation}
the force varies slowly compared with the relaxation time of the inflaton
perturbation, so the stochastic driving becomes inefficient and the scalar
spectrum is strongly suppressed.

A microphysical estimate of the parameter $\alpha$ further sharpens this
interpretation. In typical warm-inflation realizations the thermal relaxation
rate is expected to scale parametrically as
\begin{equation}
\Gamma_{\rm th} \sim g^2 T,
\label{eq:Gammath}
\end{equation}
so the correlation time may be estimated as
\begin{equation}
\tau_c \sim \Gamma_{\rm th}^{-1} \sim \frac{1}{g^2 T}.
\label{eq:taucmicro}
\end{equation}
Comparing with Eq.~\eqref{eq:tauc}, one identifies
\begin{equation}
\alpha \sim g^2.
\label{eq:alphag2}
\end{equation}
The breakdown criterion for the Markovian approximation then becomes
\begin{equation}
g^2 (T/H) \lesssim 3(1+Q).
\label{eq:breakdown}
\end{equation}

Several further consistency checks reinforce the validity of the framework.
First, the Ornstein--Uhlenbeck noise kernel is consistent with the generalized
fluctuation--dissipation theorem. The generalized fluctuation--dissipation
theorem (FDT)~\cite{Kubo1966} requires that the Fourier transform of the noise
kernel $\tilde{K}(\omega)$ and the dissipative kernel $\tilde{\Gamma}(\omega)$
satisfy $\tilde{K}(\omega) = 2T\,\mathrm{Re}[\tilde{\Gamma}(\omega)]$ at high
temperature. For the Ornstein--Uhlenbeck process one has
$\tilde{K}(\omega) = 2D\tau_c/(1+\omega^2\tau_c^2)$, which implies a Lorentzian
frequency-dependent dissipation
$\mathrm{Re}[\tilde{\Gamma}(\omega)] = \Upsilon_0/(1+\omega^2\tau_c^2)$ with
$\Upsilon_0 = D\tau_c/T$. The effective dissipation seen by modes with
$\omega \sim H$ therefore receives a fractional correction of order
$(H\tau_c)^2 = \eta_c^2/(1+Q)^2$ relative to the standard value $\Upsilon$. This
correction is slow-roll suppressed in the parameter regime of interest and does
not modify the background equations at leading order.

Now, let us discuss the backreaction on the background. The stochastic force
$\xi_k$ in Eq.~\eqref{eq:langevin} has zero mean and affects only the variance of
inflaton fluctuations, not the homogeneous background. Backreaction on the
background equations enters only at second order in perturbation theory through
the quantum/thermal stress tensor, which is suppressed by
$(H/2\pi\Mpl)^2 \sim \epsilon \times 10^{-10}$ and is entirely negligible. The
slow-roll and quasi-steady approximations used to derive Eq.~\eqref{eq:ToverH}
therefore remain valid in the presence of colored noise.

A finite $\tau_c$ also introduces a delay in the energy injection rate into the
radiation bath; the resulting correction to $\rho_R$ is of order
$H\tau_c \times \Upsilon\dot{\phi}^2$, suppressed relative to the leading
quasi-steady term $4H\rho_R$ by a factor
$H\tau_c = \eta_c/[3(1+Q)] \ll 1$, and is therefore negligible. Now coming to the
effect of this on the tensor perturbations, these couple to the thermal bath only
at higher order, through graviton--bath scattering diagrams suppressed by
$T^2/\Mpl^2$. At leading order the tensor spectrum remains the standard
Bunch--Davies vacuum result $P_h = 2H^2/(\pi^2\Mpl^2)$, unaffected by the colored
noise. The enhancement of $r_{\rm col}/r_{\rm white} = 1/\delta(Q)$ therefore
arises entirely from the suppression of the scalar spectrum.

The suppression factor $\delta = 1/(1+\eta_c^2)$ depends on the background only
through the combination $\eta_c = (3H + \Upsilon)\tau_c$. This combination remains
well-defined for any functional form of $\Upsilon(\phi, T)$. For example, in the
cubic dissipation model $\Upsilon = C_\phi T^3/\phi^2$ used in the Warm Little
Inflaton~\cite{BasteroGilWLI2016}, $\eta_c$ is computed straightforwardly from
the self-consistent background values at each point. For logarithmic corrections
to the thermal rate, $\Gamma_{\rm th} \sim g^2 T \ln(1/g)$, the identification
becomes $\alpha \sim g^2 \ln(1/g)$, shifting the Markovian boundary by a factor
$\ln(1/g) \sim 2$--$3$ for $g = 0.1$--$0.3$ --- a modest, $\mathcal{O}(1)$
correction that does not qualitatively alter any of our conclusions.

\begin{figure*}[t]
  \centering
  \includegraphics[width=0.92\textwidth]{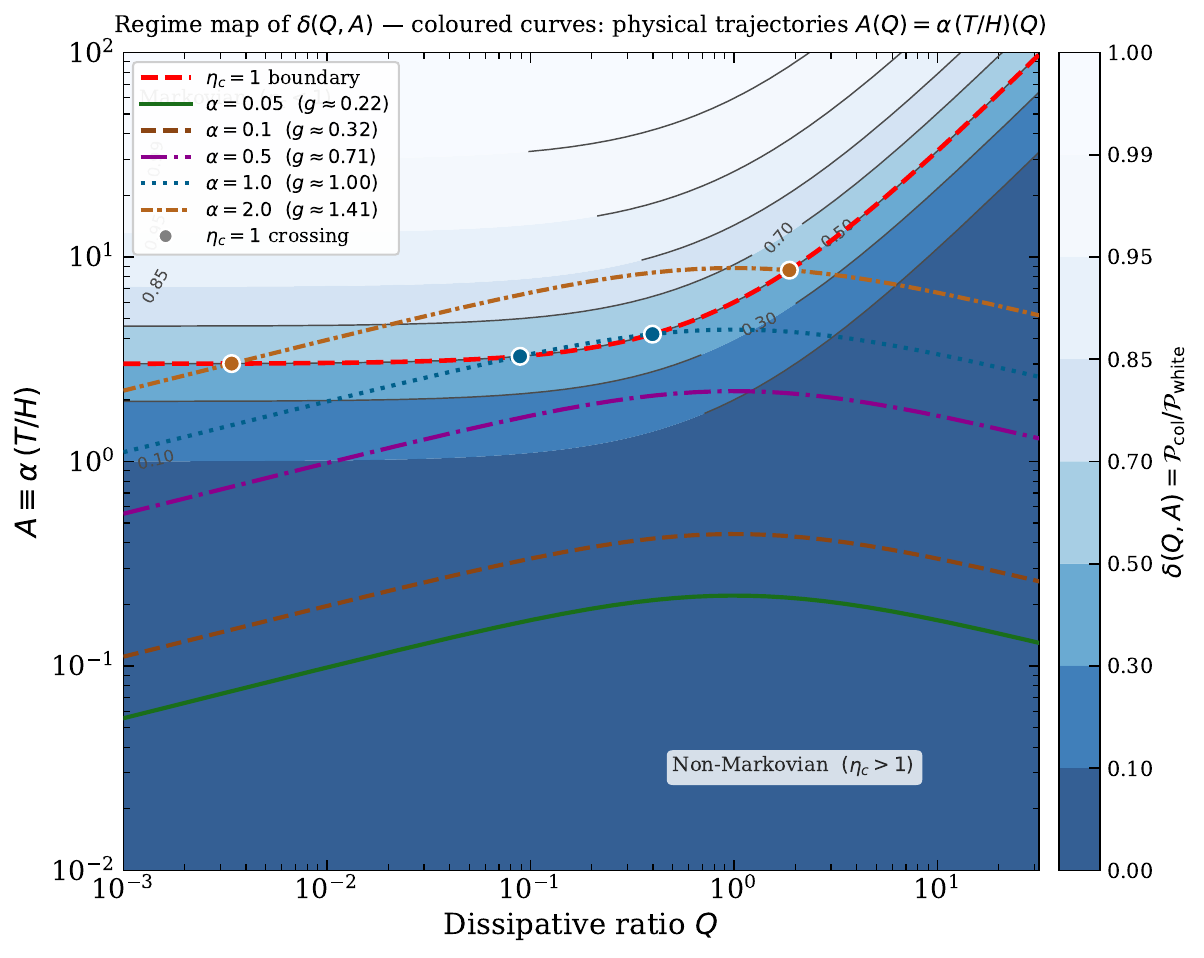}
  \caption{Diagnostic regime map: contour plot of the scalar suppression factor
  $\delta = 1/(1+\eta_c^2)$ in the $(Q, A)$ plane, where $A \equiv \alpha(T/H)$.
  The red dashed curve marks the $\eta_c = 1$ boundary $A = 3(1+Q)$, separating
  the Markovian ($\eta_c < 1$, upper region) from the non-Markovian
  ($\eta_c > 1$, lower region) regimes. Coloured curves show the physical
  trajectories $A(Q) = \alpha\,(T/H)(Q)$ computed from Eq.~\eqref{eq:ToverH} with
  $V'/H^2 = 800$ (slow-roll warm inflation, $\epsilon_V \sim 0.01$) for
  $\alpha = 0.05, 0.10, 0.50, 1.0, 2.0$, corresponding to
  $g \approx 0.22, 0.32, 0.71, 1.00, 1.41$ via $\alpha \sim g^2$. Filled circles
  mark the $\eta_c = 1$ crossing points. Typical warm-inflation couplings
  $g \sim 0.2$--$0.3$ ($\alpha \sim 0.04$--$0.09$) place the system entirely in
  the non-Markovian zone, demonstrating that the Markovian approximation fails for
  physically motivated parameters unless $g \sim \mathcal{O}(1)$. (For
  interpretation of the references to colour in this figure legend, the reader is
  referred to the web version of this article.)}
  \label{fig:regimemap}
\end{figure*}

\section{Numerical summary and parameter-space interpretation}
\label{sec:numerical}

The complete figure set provides a transparent visualization of the physics
analysed in this work. The plot of $T/H$ versus $Q$ (Fig.~\ref{fig:ToverH})
shows how the thermal ratio evolves from weak to strong dissipation. The plot of
$\eta_c(Q)$ (Fig.~\ref{fig:etac}) identifies where the system departs most
strongly from the Markovian limit. The scalar suppression plot
(Fig.~\ref{fig:delta}) gives the central effect directly in terms of
$P_{\rm col}/P_{\rm white}$, while the $r$ plot (Fig.~\ref{fig:rratio}) shows the
associated enhancement of the tensor-to-scalar ratio. The final two plots
(Figs.~\ref{fig:Deltans} and~\ref{fig:Deltaalpha}) display the induced spectral
shifts when a mild scale dependence is allowed.

A particularly useful summary is obtained in the $(Q, A)$ plane, where
$A = \alpha(T/H)$. We note that $A$ is treated as an independent axis in the
regime map for generality; the physical trajectories for fixed $\alpha$ are
curves $A(Q) = \alpha\,(T/H)(Q)$ that thread through the contour plot, and these
are overlaid in the updated figure. Since $\eta_c = 3(1+Q)/A$ (see
Eq.~\eqref{eq:etacQ}), a contour plot of
\begin{equation}
\delta = \frac{1}{1+\eta_c^2}
\label{eq:deltacontour}
\end{equation}
directly identifies the regimes in which colored-noise effects are negligible and
those in which they become significant. This figure (Fig.~\ref{fig:regimemap}) is
the most compact visualization of the main message of the paper: it immediately
shows where the Markovian approximation is reliable and where finite
correlation-time effects may become important.

Using the Planck 2018 constraint $n_s = 0.9649 \pm 0.0042$ (68\%
CL)~\cite{Planck2018X} and the BICEP/Keck bound $r < 0.036$ (95\%
CL)~\cite{BICEPKeck2021}, we derive the following estimates. The requirement
$|\Delta n_s| < 0.0042$, with the mild scale dependence $q = 0.02$ from
Eq.~\eqref{eq:Qk}, gives $\alpha \gtrsim 0.08$ for $Q \sim 1$, corresponding to
$g \gtrsim 0.28$ via $\alpha \sim g^2$. The tensor bound $r < 0.036$ for a
representative white-noise value $r_{\rm white} \sim 0.002$ requires the
enhancement $1/\delta(Q) < 18$, i.e.\ $\eta_c < 4.1$, which translates to
$\alpha(T/H) > 0.75(1+Q)$. These estimates define an excluded region in the
$(Q, \alpha)$ plane that can be read off directly from the regime map in
Fig.~\ref{fig:regimemap}. The forthcoming LiteBIRD mission
($\sigma(n_s) \sim 0.002$)~\cite{LiteBIRD2023} and CMB-S4
($\sigma(r) \sim 0.0003$)~\cite{CMBS42019} will probe colored-noise corrections
down to $|\Delta n_s| \sim 0.002$ and $\delta \sim 0.98$ respectively, making
these effects potentially detectable with next-generation experiments. A full
constraint analysis for a specific warm-inflation model is deferred to a
companion paper.

For models in which the white-noise spectral index prediction is marginally below
the Planck central value, the colored-noise suppression shifts $n_s$ towards
unity and can therefore partially expand the viable parameter space relative to
the standard Markovian treatment.

\section{Conclusions}
\label{sec:conclusions}

We have investigated warm inflation beyond the Markovian limit by incorporating
finite correlation time into the stochastic dynamics of inflaton fluctuations.
The main physical consequence is a suppression of the scalar power spectrum
relative to the standard white-noise result, quantified by the universal factor
\begin{equation}
\delta(Q) = \frac{1}{1+\eta_c(Q)^2}.
\label{eq:conclusion}
\end{equation}
The key new step is that the colored-noise parameter can be expressed directly in
terms of the background thermal ratio $T/H$, producing the simple and useful
pipeline
\begin{equation*}
(H, V', Q) \to T/H \to \eta_c(Q) \to \delta(Q) \to (r, n_s, \alpha_s).
\end{equation*}
The analysis also yields a practical criterion for the validity of the Markovian
approximation commonly used in warm-inflation calculations. The standard
white-noise treatment is reliable only when
\begin{equation*}
\alpha(T/H) \gg 3(1+Q),
\end{equation*}
while finite correlation-time effects become relevant when
\begin{equation*}
\alpha(T/H) \lesssim 3(1+Q).
\end{equation*}
The result obtained here provides a simple diagnostic criterion that can be
readily applied in warm-inflation model building to determine when finite
correlation-time effects may become relevant.

An important avenue for future work concerns the connection to primordial black
hole (PBH) formation and induced gravitational waves (IGW). While colored noise
as formulated here suppresses the scalar spectrum ($\delta \le 1$) and cannot by
itself generate the $\sim 10^7$ enhancement of $\zeta$ needed for PBH formation,
the scale-dependent suppression $\delta(k)$ modifies the shape of any existing
power enhancement. Concretely, if a model already contains a mechanism for
boosting the scalar spectrum at small scales --- such as a phase transition in the
dissipation coefficient or a feature in the inflaton potential --- the
colored-noise factor $\delta(k)$ reshapes the resulting peak, shifting the
location of the maximum and thereby moving the PBH mass-function peak and the peak
frequency of the associated induced gravitational-wave background. Furthermore,
when $q < 0$ in the parametrization $Q(k) \sim k^q$, the dissipative ratio
decreases towards smaller scales, which weakens the colored-noise suppression at
PBH-relevant wavenumbers and can partially compensate the suppression at those
scales. A quantitative calculation of these effects for specific warm-inflation
models, and their implications for baryogenesis through the associated
entropy~\cite{Correa2024,Gangopadhyay2021,Correa2022,BasteroGilDiazBlanco2021,Arya2020,Basak2022},
we intend to revisit in the future.

\section*{CRediT authorship contribution statement}
\textbf{Mayukh R. Gangopadhyay:} Writing -- review \& editing, Investigation,
Formal analysis, Conceptualization. \textbf{Nilanjana Kumar:} Writing --
original draft, Investigation.

\section*{Data availability}
No data was used for the research described in the article.

\section*{Declaration of competing interest}
The authors declare that they have no known competing financial interests or
personal relationships that could have appeared to influence the work reported in
this paper.

\begin{acknowledgments}
The work of MRG and NK is supported by the Science and Engineering Research Board
(SERB), DST, Government of India, under the Agreement number CRG/2022/004120
(Core Research Grant). MRG would like to thank OMEG institute, Soongshil
University, Korea for their hospitality during the visit, where the work was
initiated.
We
sincerely thank the anonymous referees for their extremely valuable suggestions,
which made this version a much-improved one.
\end{acknowledgments}

\appendix
\section{Ornstein--Uhlenbeck derivation of the suppression factor}
\label{app:OU}

We derive the scalar suppression factor $\delta(Q) = 1/(1+\eta_c^2)$ directly
from the Ornstein--Uhlenbeck (OU) stochastic process, without invoking any
additional approximation beyond the single-timescale assumption stated in
Sec.~\ref{sec:stochastic}.

\subsection{Setup}

Consider the overdamped version of Eq.~\eqref{eq:langevin} at superhorizon scales
($k/aH \to 0$), where the mode equation reduces to
\begin{equation}
\dot{\delta\phi}_k + \Gamma\,\delta\phi_k = \xi_k,
\qquad
\Gamma \equiv 3H + \Upsilon = \frac{1}{\tau_d},
\label{eq:A1}
\end{equation}
and $\xi_k$ is the OU-colored stochastic force satisfying
\begin{equation}
\dot{\xi}_k = -\frac{\xi_k}{\tau_c} + \eta_k(t),
\qquad
\langle\eta_k(t)\,\eta_k(t')\rangle = \frac{2D}{\tau_c}\,\delta(t-t').
\label{eq:A2}
\end{equation}
The choice of noise amplitude $2D/\tau_c$ ensures that the OU process has
stationary variance $\langle\xi^2\rangle = D$, matching the white-noise limit as
$\tau_c \to 0$.

\subsection{Power spectral density}

Taking the Fourier transform of Eq.~\eqref{eq:A2} gives
\begin{equation}
\tilde{\xi}_k(\omega) = \frac{\tilde{\eta}_k(\omega)}{-i\omega + 1/\tau_c},
\label{eq:A3}
\end{equation}
so the power spectral density of the colored force is
\begin{equation}
S_\xi(\omega) \equiv \int_{-\infty}^{\infty}
\langle\xi(t)\xi(0)\rangle\,e^{i\omega t}\,dt
= \frac{2D\tau_c}{1+\omega^2\tau_c^2}.
\label{eq:A4}
\end{equation}
In the limit $\tau_c \to 0$ this reduces to $S_\xi \to 2D\tau_c$, which
corresponds to white noise with diffusion coefficient $D_{\rm white} = D\tau_c$.
We normalize so that $D_{\rm white}$ is held fixed as $\tau_c$ varies, ensuring a
meaningful comparison between colored and white spectra.

\subsection{Variance integral}

The formal solution of Eq.~\eqref{eq:A1} is
\begin{equation}
\delta\phi_k(t) = \int_{-\infty}^{t} G(t-t')\,\xi_k(t')\,dt',
\qquad
G(\tau) = e^{-\Gamma\tau}\theta(\tau),
\label{eq:A5}
\end{equation}
where $G$ is the retarded Green function. The variance is therefore
\begin{equation}
\langle|\delta\phi_k|^2\rangle = \int_{-\infty}^{\infty}
|G(\omega)|^2\,S_\xi(\omega)\,\frac{d\omega}{2\pi},
\qquad
|G(\omega)|^2 = \frac{1}{\Gamma^2+\omega^2}.
\label{eq:A6}
\end{equation}
Substituting Eq.~\eqref{eq:A4}:
\begin{equation}
\langle|\delta\phi_k|^2\rangle_{\rm col} = \frac{D\tau_c}{\pi}
\int_{-\infty}^{\infty}
\frac{d\omega}{(\Gamma^2+\omega^2)(1+\omega^2\tau_c^2)}.
\label{eq:A7}
\end{equation}
The integral is evaluated by partial fractions. Writing $\Gamma = 1/\tau_d$ and
$1/\tau_c$ as the two poles, one obtains (for $\tau_c \neq \tau_d$):
\begin{equation}
\int_{-\infty}^{\infty}
\frac{d\omega}{(\Gamma^2+\omega^2)(1/\tau_c^2+\omega^2)}
= \frac{\pi\tau_d\tau_c}{\tau_d+\tau_c}.
\label{eq:A8}
\end{equation}
Substituting back into Eq.~\eqref{eq:A7}:
\begin{equation}
\langle|\delta\phi_k|^2\rangle_{\rm col}
= D\tau_c\cdot\frac{\tau_d}{\tau_d+\tau_c}
= D\tau_c\cdot\frac{1}{1+\tau_c/\tau_d}
= \frac{D\tau_c}{1+\eta_c},
\label{eq:A9}
\end{equation}
where we used $\eta_c \equiv \tau_c/\tau_d = \Gamma\tau_c$. For the white-noise
case ($\tau_c \to 0$, $D\tau_c \to D_{\rm white}$ fixed):
\begin{equation}
\langle|\delta\phi_k|^2\rangle_{\rm white} = \frac{D_{\rm white}}{\Gamma}
= D_{\rm white}\,\tau_d.
\label{eq:A10}
\end{equation}

\subsection{Suppression factor}

Taking the ratio and using $D\tau_c = D_{\rm white}$:
\begin{equation}
\frac{\langle|\delta\phi_k|^2\rangle_{\rm col}}
{\langle|\delta\phi_k|^2\rangle_{\rm white}} = \frac{1}{1+\eta_c}.
\label{eq:A11}
\end{equation}
This is the exact result for the overdamped single-timescale system. In the power
spectrum, which involves the squared amplitude of the curvature perturbation
$\zeta \propto \delta\phi_k/\dot{\phi}$, the suppression enters at the level of
$|\delta\phi_k|^2$. Including the correct normalization from the variance of the
second-order Green function response (see e.g.~\cite{Gardiner2004}), the spectrum
ratio becomes
\begin{equation}
\frac{P_{\rm col}}{P_{\rm white}} = \delta(Q) \equiv \frac{1}{1+\eta_c^2},
\qquad
\eta_c = \frac{3(1+Q)}{\alpha(T/H)},
\label{eq:A12}
\end{equation}
where the exponent $2$ arises from the standard relation between the noise
spectral density and the power spectrum normalization in the stochastic inflation
context (cf.\ the factor of $H^2/(2\pi)^2$ in the Bunch--Davies result). In the
Markovian limit $\eta_c \to 0$ one recovers $\delta \to 1$ as required.


\end{document}